\documentclass{ws-ijmpe}
\usepackage[super]{cite}
\usepackage{xcolor}
\usepackage{graphicx}
\usepackage{amsmath}
\usepackage{algorithm}
\usepackage{algpseudocode}
\usepackage{booktabs}
\usepackage{comment}
\usepackage{svg}

\usepackage[verbose,hypertexnames=false]{hyperref}
\hypersetup{colorlinks=false,allbordercolors=blue,pdfborderstyle={/S/U/W 1}}
\begin{document}

\markboth{Anwar Ibrahim et al.}{Reinforcement Learning for Accelerator Beamline Control: a simulation-based approach}

\begin{comment}
\catchline{}{}{}{}{}
\end{comment}
%%%%%%%%%%%%%%%%%%%%%%%%%%%%%%%%%%%%%%%%%%%%%%%%%%%%%%%%%%%%%%%%%%%%

\title{Reinforcement Learning for Accelerator Beamline Control: a simulation-based approach}

\author{Anwar Ibrahim}

\address{Laboratory of Methods for Big Data Analysis, HSE University, 11 Pokrovsky Boulevard, Moscow, 109028, Russia\\
aibrahim@hse.ru}

\author{Alexey Petrenko}

\address{Budker Institute of Nuclear Physics, 11, Akad. Lavrent'eva Prospekt, Novosibirsk, 630090, Russia\\ A.V.Petrenko@inp.nsk.su}

\author{Maxim Kaledin}

\address{Laboratory of Methods for Big Data Analysis,HSE University,  11 Pokrovsky Boulevard, Moscow, 109028, Russia\\
mkaledin@hse.ru}

\author{Ehab Suleiman}

\address{Tikhonov Moscow Institute of Electronics and Mathematics,HSE University,  34 Tallinskaya Ulitsa, Moscow, 123458, Russia\\
e.suleiman@hse.ru}

\author{Fedor Ratnikov}

\address{Laboratory of Methods for Big Data Analysis,HSE University,  11 Pokrovsky Boulevard, Moscow, 109028, Russia\\
fratnikov@hse.ru}

\author{Denis Derkach}

\address{Laboratory of Methods for Big Data Analysis,HSE University,  11 Pokrovsky Boulevard, Moscow, 109028, Russia\\
dderkach@hse.ru}

\maketitle

\begin{comment}
    
\begin{history}
\received{Day Month Year}
\revised{Day Month Year}
\accepted{Day Month Year}
\published{Day Month Year}
\end{history}
\end{comment}

\begin{abstract}
Particle accelerators play a pivotal role in advancing scientific research, yet optimizing beamline configurations to maximize particle transmission remains a labor-intensive task requiring expert intervention. In this work, we introduce RLABC (Reinforcement Learning for Accelerator Beamline Control), a Python-based library that reframes beamline optimization as a reinforcement learning (RL) problem. Leveraging the Elegant simulation framework, RLABC automates the creation of an RL environment from standard lattice and element input files, enabling sequential tuning of magnets to minimize particle losses. We define a comprehensive state representation capturing beam statistics, actions for adjusting magnet parameters, and a reward function focused on transmission efficiency. Employing the Deep Deterministic Policy Gradient (DDPG) algorithm, we demonstrate RLABC's efficacy on two beamlines, achieving transmission rates of 94\% and 91\%, comparable to expert manual optimizations. This approach bridges accelerator physics and machine learning, offering a versatile tool for physicists and RL researchers alike to streamline beamline tuning.
\end{abstract}


\ccode{PACS Number(s): 35J10}

\section{Introduction}
\label{sec:intro}

Particle accelerators are indispensable tools in modern science, enabling breakthroughs in high-energy physics, materials science, and medical applications such as cancer therapy~\cite{lee2020, chao2013}. These complex systems use electromagnetic fields to propel charged particles and typically consist of a particle source and a carefully engineered \textbf{beamline} to guide the particles from the source to a target, such as an experimental chamber where valuable data is collected~\cite{wiedemann2015}. Beamlines utilize quadrupole magnets to focus or defocus the particle beam and dipole magnets to steer it, ensuring precise delivery to the target. However, optimizing magnet settings to maximize particle transmission remains a significant challenge, as it relies solely on magnetic field adjustments. This process demands considerable time and expertise due to the high-dimensional parameter space and dynamic beam behavior~\cite{wolski2014}.

Traditional beam optimization methods, such as manual tuning or simplex algorithms, are time-intensive and require substantial expertise to achieve optimal performance~\cite{nelder1965}. To address these limitations, we introduce RLABC (Reinforcement Learning for Accelerator Beamline Control), a novel Python-based library that reformulates beamline optimization as a reinforcement learning (RL) problem~\cite{sutton2018, mnih2015}. RL has demonstrated effectiveness in tackling complex optimization tasks with high-dimensional spaces and dynamic environments, as evidenced in robotics~\cite{singh2022reinforcement}, games (e.g., Atari, chess, and Go)~\cite{silver2018general}, autonomous driving~\cite{kiran2022deep}, and large language models~\cite{deepseek2025deepseekr1}. For beamline optimization, RL is particularly well-suited due to its capability for sequential decision-making under uncertainty, facilitating efficient exploration of vast parameter spaces. 

RLABC leverages the \textbf{Elegant} simulation software and SDDS toolkit to model beams and beamlines, using \texttt{.lte} (lattice) and \texttt{.ele} (element) files as inputs~\cite{borland2000}. From these \textbf{Elegant} input files, RLABC dynamically generates a fully functional RL environment without requiring additional configuration or physicist intervention, thereby streamlining the optimization workflow. The primary goal of the environment to maximize particle transmission through the beamline by fine-tuning magnet parameters. This RL environment provides a versatile platform for both RL researchers and accelerator physicists. RL experts can integrate custom algorithms into RLABC or utilize those from the \textit{Stable-Baselines3} library to develop tailored solutions for beamline challenges~\cite{raffin2021}. Meanwhile, physicists without programming expertise can leverage RLABC's built-in algorithms to optimize beamlines effortlessly. Unlike libraries designed for specific beamlines, RLABC requires only standard \textbf{Elegant} input files, automating the tuning process and bridging accelerator physics with machine learning.

\section{Theoretical Background}
\label{sec:Theory}

\subsection{Reinforcement Learning Basics}

Reinforcement Learning (RL) is a branch of machine learning where an agent learns to make sequential decisions by interacting with an environment through trial and error, without relying on labeled or unlabeled data. This approach is has its success stories in solving real-world problems, such as autonomous driving by Wayve \cite{kendall2018learning}, robotics \cite{ibarz2024deep}, financial trading \cite{li2024review}, healthcare optimization \cite{liu2023deep}.

To define an RL problem, we need to define an \textbf{Agent}, the intelligent unit that is responsible for making decisions, and that can be trained to achieve an objective, and an \textbf{environment}, the setting where the agent operates. the environment is defined by correctly defining these concepts:

\begin{itemize}
    \item \textit{States ($S_t$)}: Environmental overviews at moment $t$, supplying the agent with data for upcoming decisions; the following state is $S_{t+1}$.
    \item \textit{Actions ($A_t$)}: Options the agent uses to alter the setting at $t$.
    \item \textit{Rewards ($R_t$)}: Numerical signals from the environment at $t$, promoting favorable outcomes and discouraging poor ones; next is $R_{t+1}$.
\end{itemize}

To correctly define the RL problem, it must satisfy the Markov property, meaning the \textit{next state} ($S_{t+1})$ and \textit{reward} ($R_{t+1}$) are dependent only on the \textit{current state} ($S_t$) and \textit{action} ($A_t$), not on prior history, enabling the problem to be modeled as a Markov Decision Process (MDP).

\begin{figure}[h]
    \centering
    \includegraphics[width=0.8\textwidth]{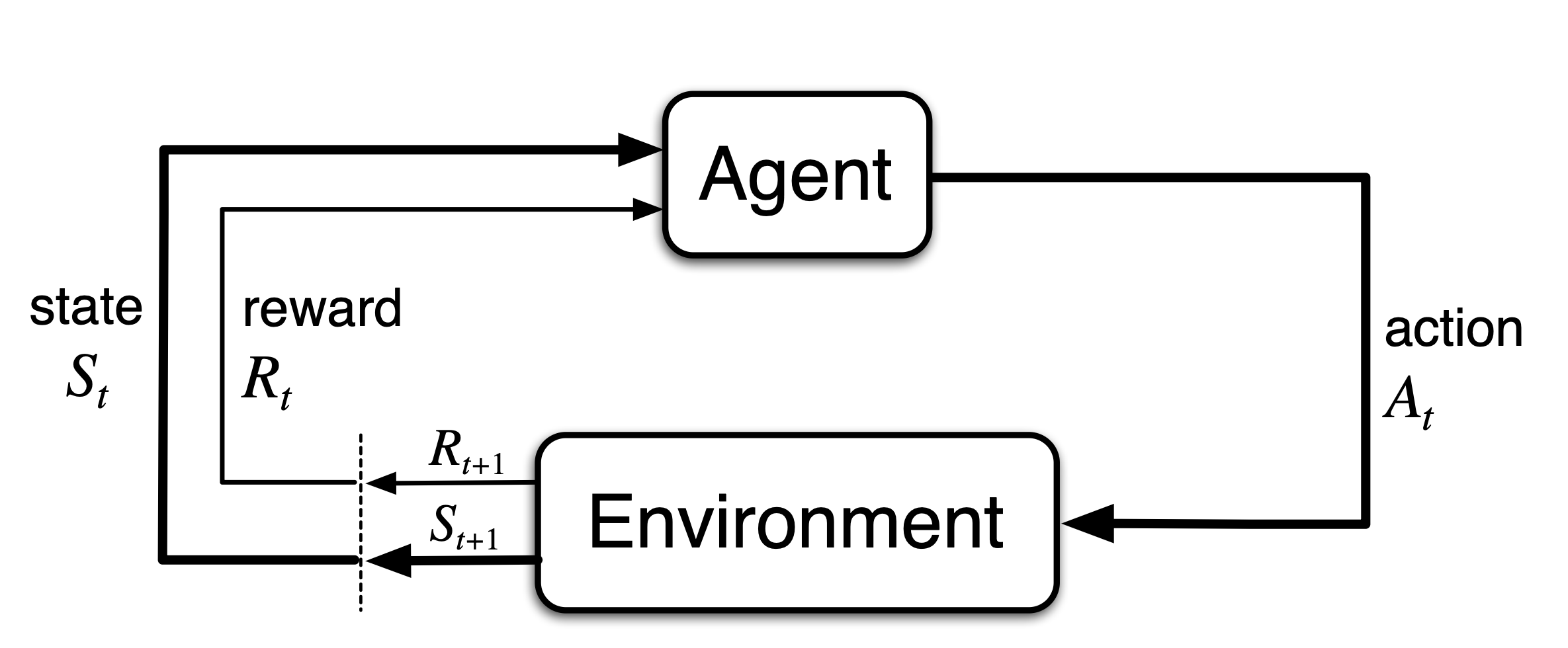}
    \caption{RL interaction cycle: At step $t$, the agent reviews state $S_t$ and reward $R_t$, picks action $A_t$, moves to $S_{t+1}$ with $R_{t+1}$, and iterates to refine performance.}
    \label{fig:rl_loop}
\end{figure}

RL uses trail and errors to explore the action space, it also balances the explorations and exploitations to maximize the return \( G_t \) which is the discounted sum of future rewards: \( G_t = \sum_{k=0}^{\infty} \gamma^k R_{t+k+1} \) where \( \gamma \in (0, 1) \) is the discount factor, balancing the importance of immediate versus future rewards.

\subsection{Simulation Framework}

In Reinforcement learning, we can use either real environment or simulation environments, and for this specific task we will use simulation environments because they are efficient, safer and cost-effective, and trying different setups on the complex and costly particle accelerators can cause serious problems or damage the equipments. There are several programs that can be used to create beam simulations like MAD-X, Astra and \textbf{Elegant}. In this work we are using \textbf{Elegant}, which is a deliberate design preference.
Elegant needs two input files to operate:
\begin{enumerate}
    \item \texttt{.lte} file, which defines the beamline design (e.g., magnets, drifts, apertures).
    \item \texttt{.ele} file, which specifies beam properties (e.g., particle distributions, energy levels, tracking parameters).
\end{enumerate}
Its outputs include \texttt{.sdds} files from watch points, providing critical beam data such as ($x$,$y$) coordinates (in meters) for each particle’s position, and ($xp$, $yp$) angles representing the particle’s momentum direction, which form the basis for state representations in our reinforcement learning framework

\section{Methodology}
\label{sec:methodology}

To formulate the beamline optimization task as a reinforcement learning (RL) problem, we follow the standard RL paradigm shown in Figure~\ref{fig:rl_loop}. This paradigm consists of an \textit{agent} that selects \textit{actions} and an \textit{environment} that executes these actions, simulates the resulting beamline state, and returns the new \textit{state}, the corresponding \textit{reward}, and a \textit{done} flag.

RL excels in sequential decision-making tasks, where decisions are made step-by-step over time. To adapt beamline optimization to this framework, we address two key challenges: (1) transforming beamline tuning into a sequential decision-making process, and (2) defining the \textit{state}, \textit{action}, \textit{reward}, and \textit{done} flag appropriately.

To address the first challenge, we define an \textit{episode} as the longitudinal progression along the beamline from start to end. In each episode, the agent observes the particle distribution at specific points, adjusts controllable elements (e.g., magnets), and advances forward, aiming to maximize particle transmission.

Defining this environment requires a faithful simulation of beamline dynamics and a mapping of the physical problem into RL terms: states, actions, and rewards. A critical aspect is ensuring the Markov property, where the next state depends only on the current state and action, not on prior history. To achieve this, the state must capture sufficient information about the particle distribution just before each decision, and the beamline must be structured to allow observations immediately preceding controllable elements.

\begin{figure}[ht]
\centering
\includegraphics[width=0.9\textwidth]{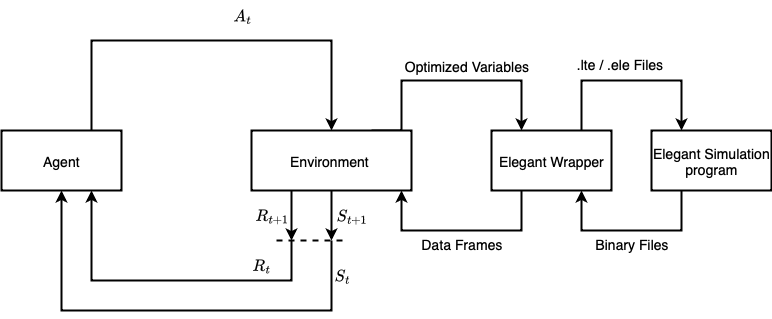}
\caption{System layout of RLABC: The RL agent interacts with the environment, which communicates with the Elegant Wrapper and simulation engine.}
\label{fig:system_layout}
\end{figure}

In our implementation, the simulation backend is provided by the \textbf{Elegant} program, which models beam dynamics based on accelerator lattice and element configurations. To interface between Python, RL libraries, and Elegant, we developed a custom \textbf{Elegant Wrapper}. This wrapper abstracts Elegant's complexity, enabling seamless communication with the RL environment and agent. The overall architecture of RLABC is illustrated in Figure~\ref{fig:system_layout}.

\subsection{Elegant Wrapper}

The Elegant Wrapper is responsible for connecting Python to the Elegant simulation software. It takes as input the \texttt{.lte} (lattice) and \texttt{.ele} (element) files and provides several key functionalities:

\begin{itemize}
\item \textbf{Beamline parsing:} Reads input files and constructs a graph-based representation of the beamline, identifying magnets, their positions, and element apertures (defined as the maximum transverse extent in $x$ and $y$ for the elliptical cross-section, describing the beamline's geometry and width constraints).
\item \textbf{Beamline Preprocessing:} modifies the beamline architecture regardless of its original configuration. It inserts watch points immediately before each optimizable element (quadrupole or dipole magnets), resulting in a standardized structure: non-optimizable elements (e.g., apertures, drifts) $\rightarrow$ watch point $\rightarrow$ optimizable element $\rightarrow$ non-optimizable elements $\rightarrow$ watch point $\rightarrow$ optimizable element, and so on. This preprocessing ensures that particle distributions can be observed just before each control decision, preserving the Markov property by making the next state dependent only on the current state and action.

\begin{figure}[h!]
    \centering
    \includegraphics[width=0.9\textwidth]{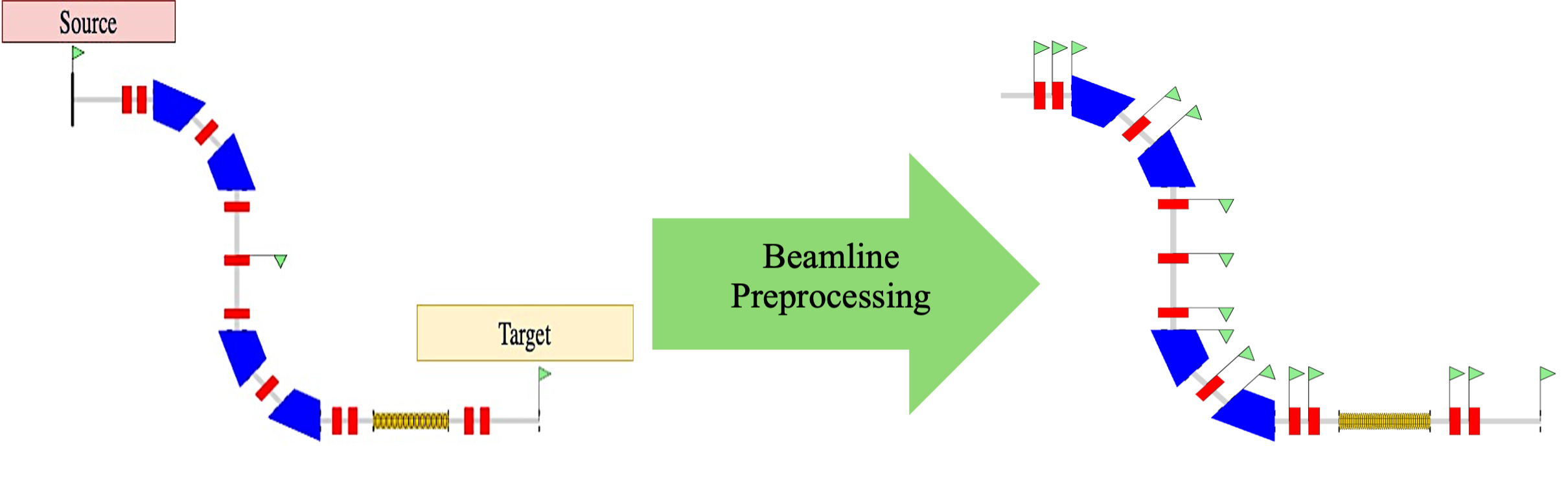}
    \caption{the beamline architecture after applying the preprocessing step. the blue blocks are Dipoles magnets, the red blocks are quadruples, the green flags are the watch points, also the yellow block is radio frequency cavities. after applying the preprocessing step we will get a watch point before every controallable element, and we will remove all other watch poins.}
    \label{fig:beamline_preprocessing}
\end{figure}

\item \textbf{Simulation control:} Allows modification of simulation parameters such as the number of initial particles or values of different variables in the \texttt{.lte} file before executing Elegant runs, then runs the Elegant simulation after.

\item \textbf{Data handling:} Parses Elegant’s outputs and converts them from binary files into Python-friendly structures such as data frames, which are directly usable by the RL environment and agent.

\item \textbf{Visualization:} Provides plotting functions for standard accelerator physics quantities
\begin{itemize}
\item Beta ($\beta$): Beam size variation.
\item Alpha ($\alpha$): Envelope slope.
\item Gamma ($\gamma$): $(1 + \alpha^2)/\beta$, divergence link.
\item Dispersion ($D$): Energy-based path shift.
\item Emittance ($\epsilon$): Phase space measure, quality indicator.
\item Tune ($\nu$): Oscillations per cycle, stability factor.
\item Chromaticity ($\xi$): Tune-energy change.
\end{itemize}
\end{itemize}

The wrapper makes it easier to interact with Elegant.

\subsection{Environment}

The environment episodes procedure in this setup begins by utilizing the Elegant Wrapper to preprocess the beamline lattice, specifically inserting watch points immediately before each optimizable element—such as magnets that can be tuned for beam control—ensuring precise monitoring of particle beam dynamics. Following this, a graph representation of the beamline is constructed to model the interconnected components and their relationships efficiently. The process then initializes the current watch point to the initial one at the start of the beamline. While the current watch point has not reached the end of the beamline, the algorithm identifies the next watch point along with the associated optimizable element as the next optimization step. It then observes the \textit{state} $S_t$ at the current watch point, which encapsulates key beam properties such as position and momentum derived from the simulation data. Based on this \textit{state}$S_t$  , an \textit{action}$A_t$  is selected for the optimizable element, representing adjustments like changes in magnet settings to influence the beam trajectory. The \textit{action} $A_t$  is applied to the element, and the Elegant simulation is executed to propagate the particle beam forward to the next watch point, updating the system's configuration. After the simulation advance, a \textit{reward} $R_{t+1}$  is computed based on how well the new beam parameters align with desired objectives "minimizing the number of lost particles"  to guide the agent's learning. Finally, the current watch point is updated to the next one, continuing the iterative process until the beamline's end is reached.

We designed our \textit{state}, \textit{reward}, \textit{action} as the following:

\textbf{State.} Each state is represented as a fixed-length feature vector (57 dimensions), derived from the simulation output at watch points.
\begin{itemize}
\item Statistical summaries (16 values): Median, interquartile range (IQR), 10th percentile, and 90th percentile for each of the transverse positions ($x$, $y$) and angular deviations ($x'$, $y'$).
\item 2D histogram (25 values): A normalized 5$\times$5 histogram of particle positions in the ($x$, $y$) plane, capturing beam shape and spread.
\item Beam statistics (1 values): Total number of particles (point count).
\item Magnet type identifier (1 value): A flag where 1 indicates a quadrupole magnet and 0 indicates a dipole magnet, informing the agent of the type being optimized.
\item Covariance structure (10 values): The upper triangle of the covariance matrix for the standardized [$x$, $y$, $x'$, $y'$] data, encoding correlations between beam coordinates.
\item Aperture parameters (4 values): Ellipse parameters before ($a_{prev}$, $b_{prev}$) and after ($a_{next}$, $b_{next}$) the optimizatble element and its watch point as ($a_{prev}$, $b_{prev}$) $\rightarrow$ watch point $\rightarrow$ element $\rightarrow$ ($a_{next}$, $b_{next}$).
\end{itemize}
This design ensures a fixed-size state unaffected by the number of beamline elements, their types, or the number of particles, facilitating generalization to different beamlines architectures, or initial particle counts.

\textbf{Action.}  
The environment has a 4-dimensional continuous action space. At each step, the agent adjusts the parameters of a single controllable element.  

For quadrupoles, the tunable variables include the focusing strength $K1$, which can vary between $-25$ and $25\,\text{m}^{-2}$, and small horizontal and vertical steering kicks (HKICK, VKICK), each constrained to the range $[-5 \times 10^{-3}, 5 \times 10^{-3}]$ radians.  

For dipoles, the agent can apply a fractional field strength error (FSE) within $[-5 \times 10^{-3}, 5 \times 10^{-3}]$, allowing subtle modifications to bending and focusing properties.  

Only the parameters relevant to the current element are activated, If the current element is a quadrupole, the first three values (K1, HKICK, VKICK) are applied, and FSE is ignored. If it is a dipole, only FSE is applied, and the first three are ignored.

\textbf{Reward.} The reward encourages maximizing particle transmission while remaining independent of the initial particle count and focusing on fractional losses per step. Let $t$ be the current step index, $N_0$ the initial number of particles, $N_t$ the particles at the current watch point, $N_{t-1}$ the particles at the previous watch point, $M$ the maximum number of controllable elements in the episode. Define the base term $B_t = N_t / N_0$. When $N_t \leq 3$, a penalty term is applied,
$P_t = \sqrt{\lvert M^2 - t^2 \rvert} / M$. The reward is then:
\[
R_t =
\begin{cases}
B_t - P_t, & \text{if } N_t \leq 3, \\
B_t \cdot \dfrac{N_t}{N_{t-1}}, & \text{if } N_t > 3~.
\end{cases}
\]
This makes the reward proportional to the fraction of surviving particles and scales it by the per-step retention rate to minimize losses at individual elements. If $N_t \leq 3$, a failure is assumed and a larger penalty is applied earlier in the beamline (higher $P_t$ for smaller $k$). The episode terminates if $N_t \leq 3$ or once the beamline end is reached. Overall, higher rewards correspond to more particles reaching the end, aligning with the optimization goal.

\subsection{Agent}

To solve the reinforcement learning (RL) problem defined by our environment, we carefully selected an algorithm tailored to its unique characteristics. Numerous RL algorithms are available for continuous control tasks, but the choice depends on factors such as the action space, the feasibility of model-based approaches, and the balance between exploration and exploitation. Our environment features a continuous action space, rendering algorithms designed for discrete actions unsuitable. Furthermore, modeling the complex dynamics of particle interactions in the beamline is a formidable challenge, leading us to favor model-free methods that learn directly from interactions with the environment. To effectively balance exploration of the parameter space with exploitation of learned policies, we chose the Deep Deterministic Policy Gradient (DDPG) algorithm\cite{lillicrap2016} . DDPG is well-suited for continuous action spaces and employs an actor-critic architecture to ensure stable and efficient learning. To enhance exploration, we incorporated Gaussian noise into the action selection process, promoting sufficient variability during training.

The hyperparameters for DDPG are presented in Table~\ref{table:ddpg_hyperparams}.

\begin{table}[h]
\centering
\small
\begin{tabular}{|l|c|}
\hline
\textbf{Hyperparameter} & \textbf{Value} \\ \hline
Actor learning rate ($\alpha$) & $1 \times 10^{-4}$ \\ \hline
Critic learning rate ($\beta$) & $1 \times 10^{-3}$ \\ \hline
Batch size & 128 \\ \hline
Discount factor ($\gamma$) & 0.99 \\ \hline
Soft update ($\tau$) & 0.005 \\ \hline
Replay buffer size & 1,000,000 \\ \hline
Noise type & Gaussian \\ \hline
\end{tabular}
\caption{Hyperparameters for the DDPG algorithm used in the experiments.}
\label{table:ddpg_hyperparams}
\end{table}

\section{Results}

\begin{figure}[h]
    \centering
    \includegraphics[width=0.8\textwidth]{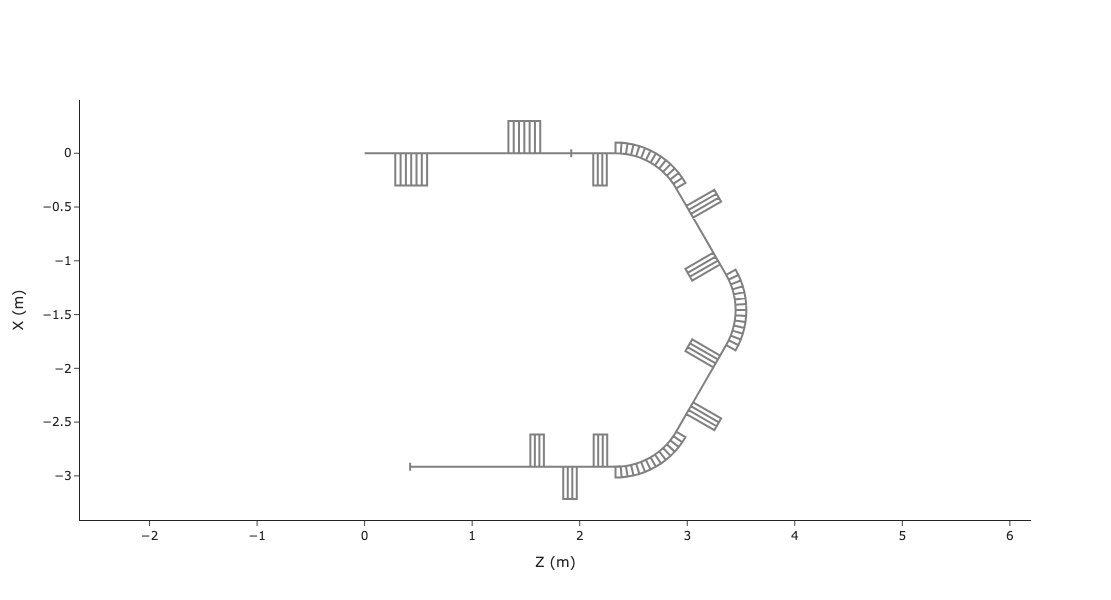}
    \caption{The architecture of the first beamline: tall triangles represent the quadrupoles, while the shorter ones that bend the beam are the dipoles.}
    \label{fig:beamline3_bends}
\end{figure}

\begin{figure}[h]
    \centering
    \includegraphics[width=0.8\textwidth]{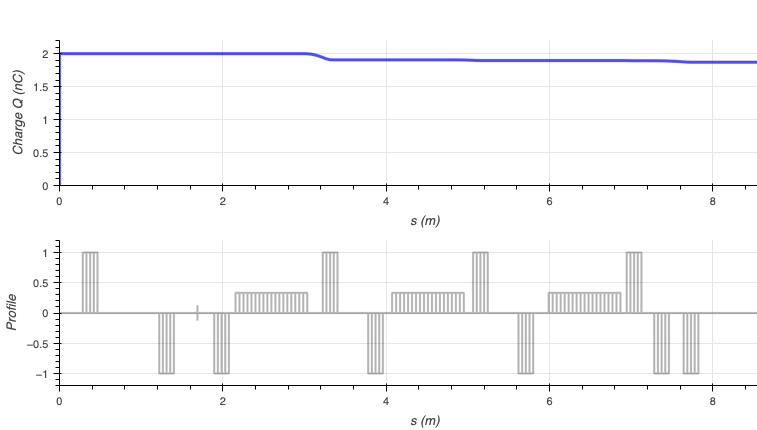}
    \caption{The beam charge along the first beamline, demonstrating minimal particle loss as the charge remains nearly constant.}
    \label{fig:charge}
\end{figure}

\begin{figure}[h]
    \centering
    \includegraphics[width=0.8\textwidth]{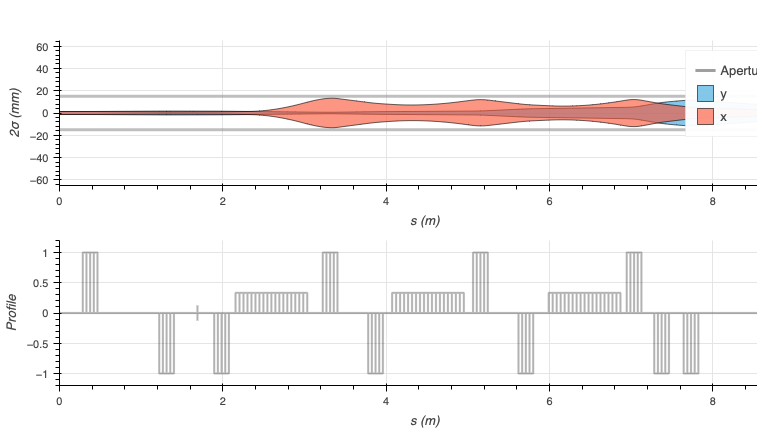}
    \caption{Beam envelope along the x-axis (red) and y-axis (blue) for the first beamline, with gray lines denoting the aperture. The beam remains fully contained within the aperture throughout.}
    \label{fig:Envelope_Aperture}
\end{figure}

\begin{figure}[h]
  \centering
  \begin{minipage}[b]{0.45\textwidth}
    \includegraphics[width=\textwidth]{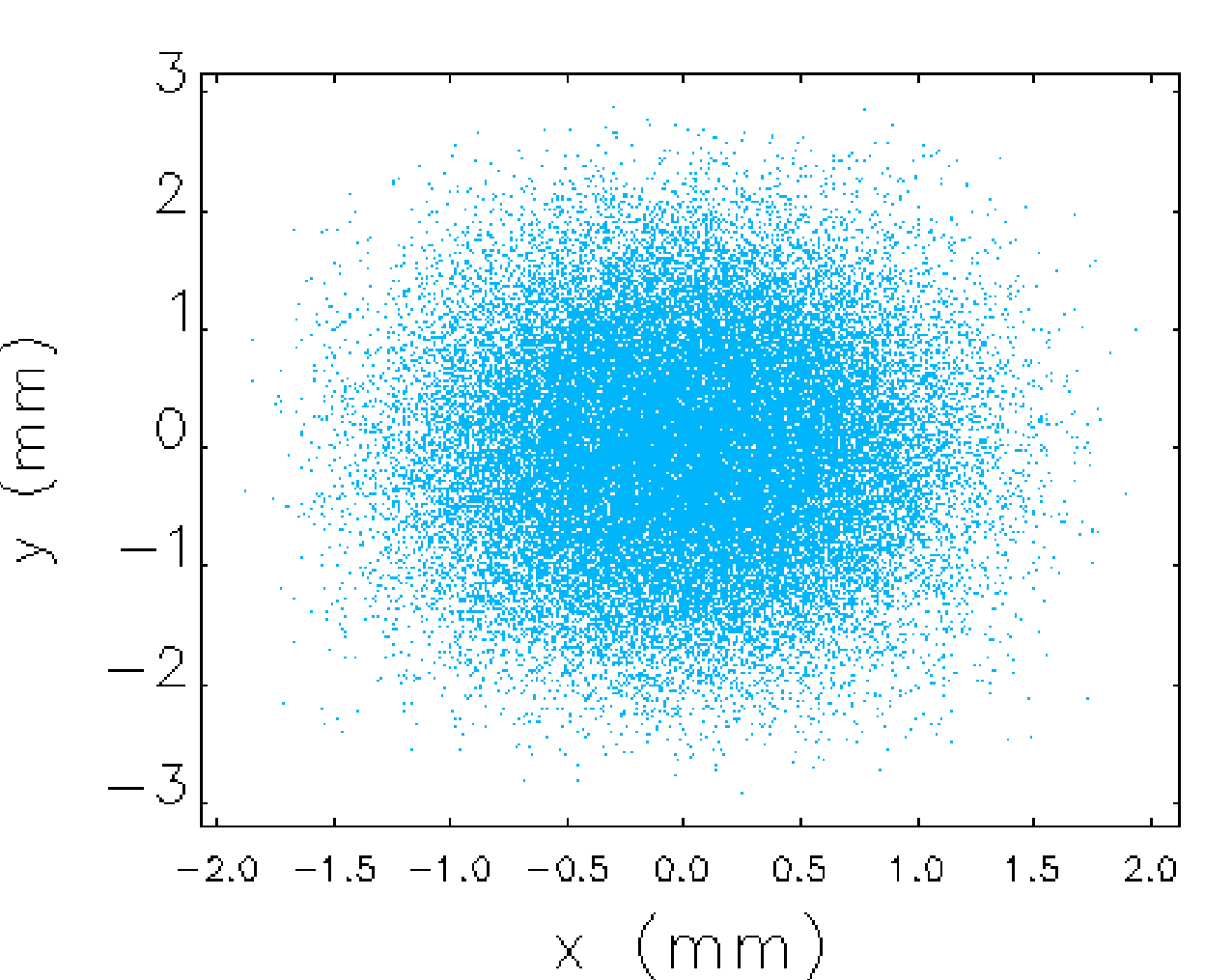}
    \caption{Particle distribution at the beginning of the first beamline.}
  \end{minipage}
  \hfill
  \begin{minipage}[b]{0.45\textwidth}
    \includegraphics[width=\textwidth]{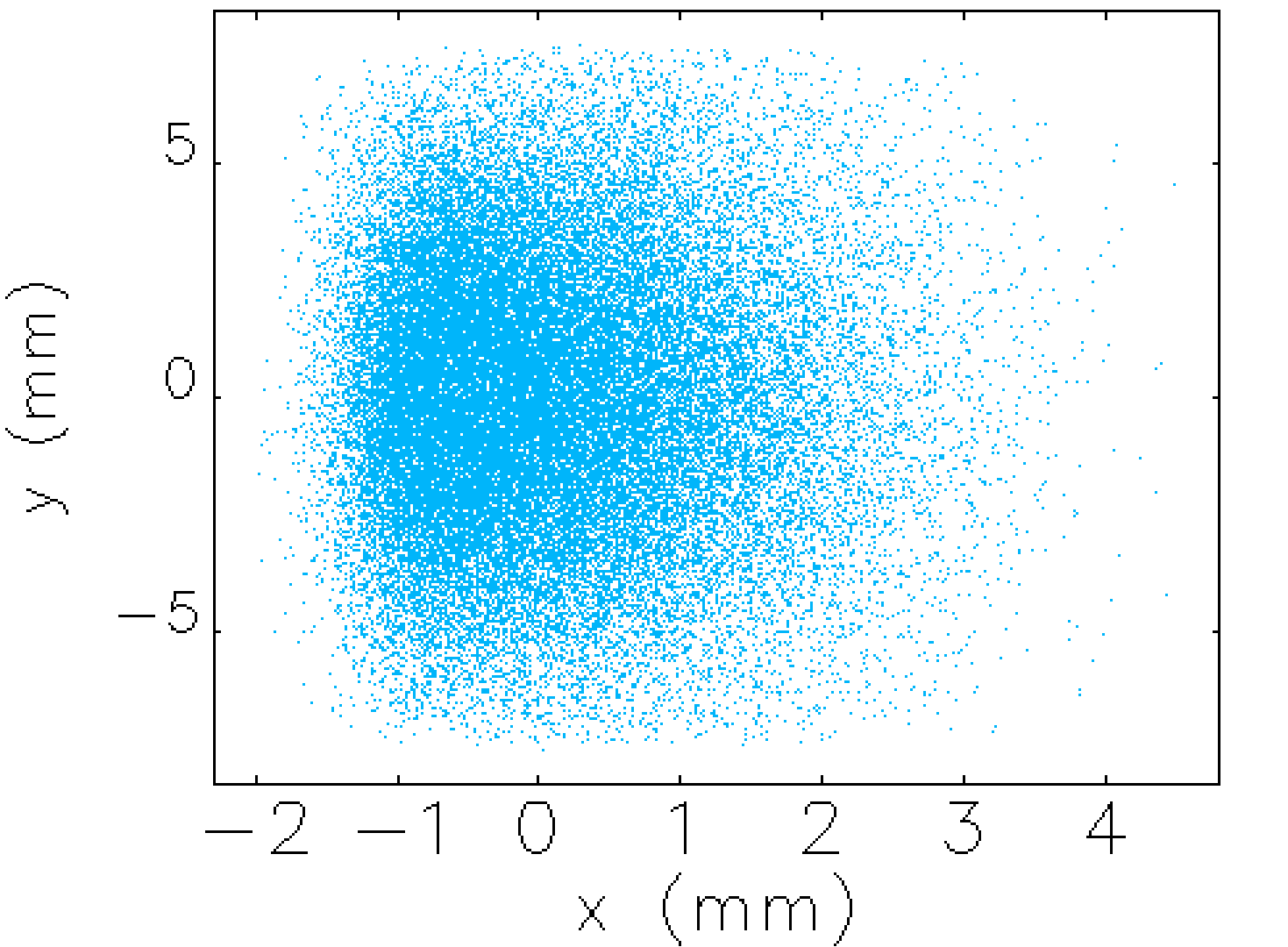}
    \caption{Particle distribution at the end of the first beamline.}
  \end{minipage}
  \label{fig:particles_dis}
\end{figure}

We evaluated our RLABC system across several beamlines presenting diverse challenges. In this section, we highlight the performance on two representative beamlines. The first beamline, illustrated in Figure~\ref{fig:beamline3_bends}, consists of 13 magnets: 10 quadrupoles for focusing and defocusing the beam, and 3 dipoles responsible for bending the particle trajectory. The primary challenge in this setup is managing the beam through three distinct bends while maintaining high transmission efficiency. After training our DDPG agent for 5000 episodes, we achieved a 94\% transmission rate, which closely aligns with the 93\% transmission obtained by our physics experts through manual optimization. Figure~\ref{fig:charge} depicts the beam charge along the beamline, where any decrease in charge indicates particle loss; notably, the charge remains nearly constant, signifying minimal losses. For a more detailed visualization of the beam envelope, Figure~\ref{fig:Envelope_Aperture} shows the beam extent along the x-axis (in red) and y-axis (in blue), with gray lines representing the aperture boundaries. The envelope stays well within the aperture throughout, confirming effective containment. Additionally, Figure~\ref{fig:particles_dis} compares the particle distribution at the beginning and end of the beamline, demonstrating that the beam remains coherent with little dispersion.

\begin{figure}[h]
    \centering
    \includegraphics[width=0.8\textwidth]{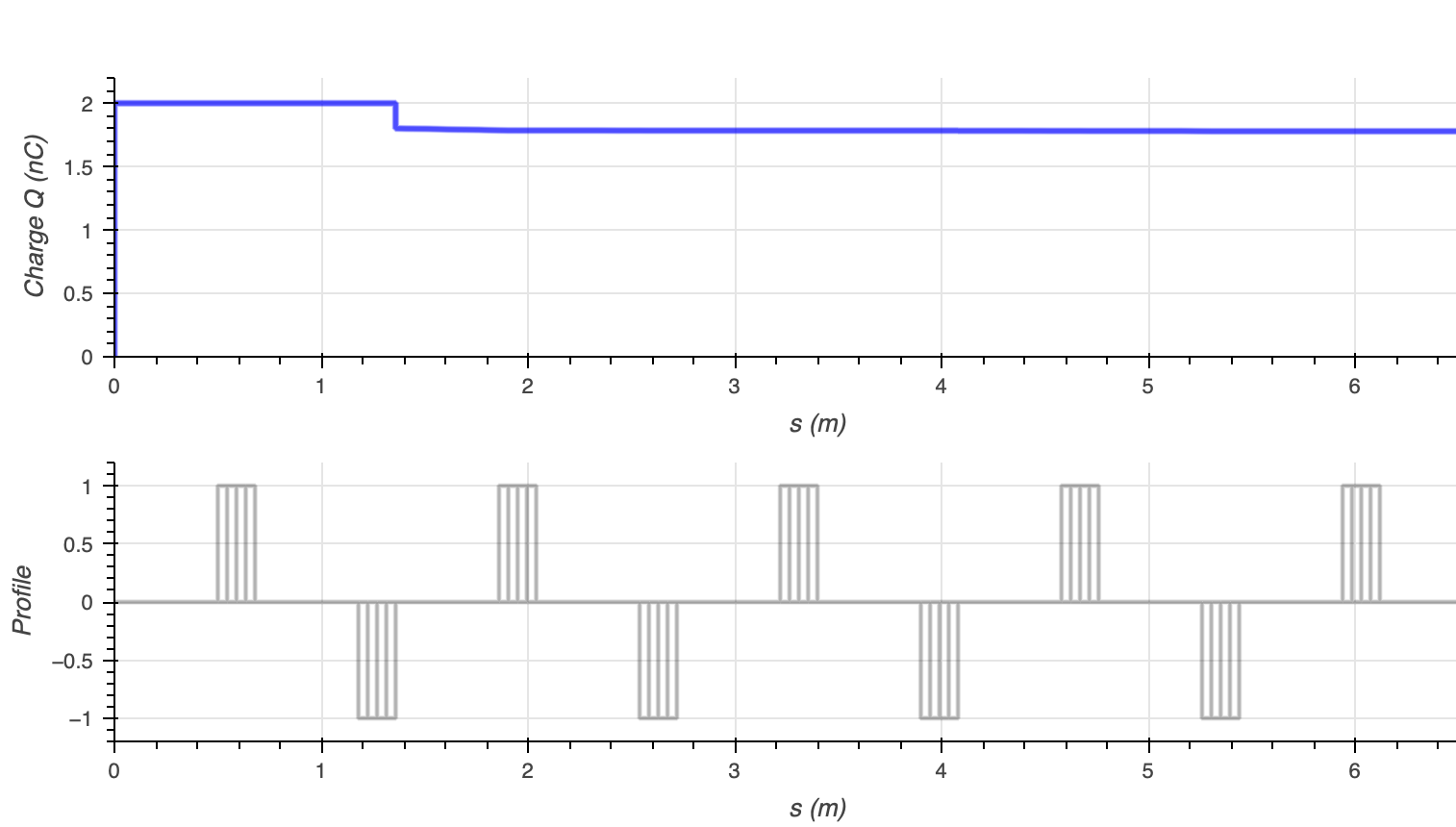}
    \caption{The beam charge along the second beamline, indicating minimal particle loss.}
    \label{fig:charge_2}
\end{figure}

\begin{figure}[h]
    \centering
    \includegraphics[width=0.8\textwidth]{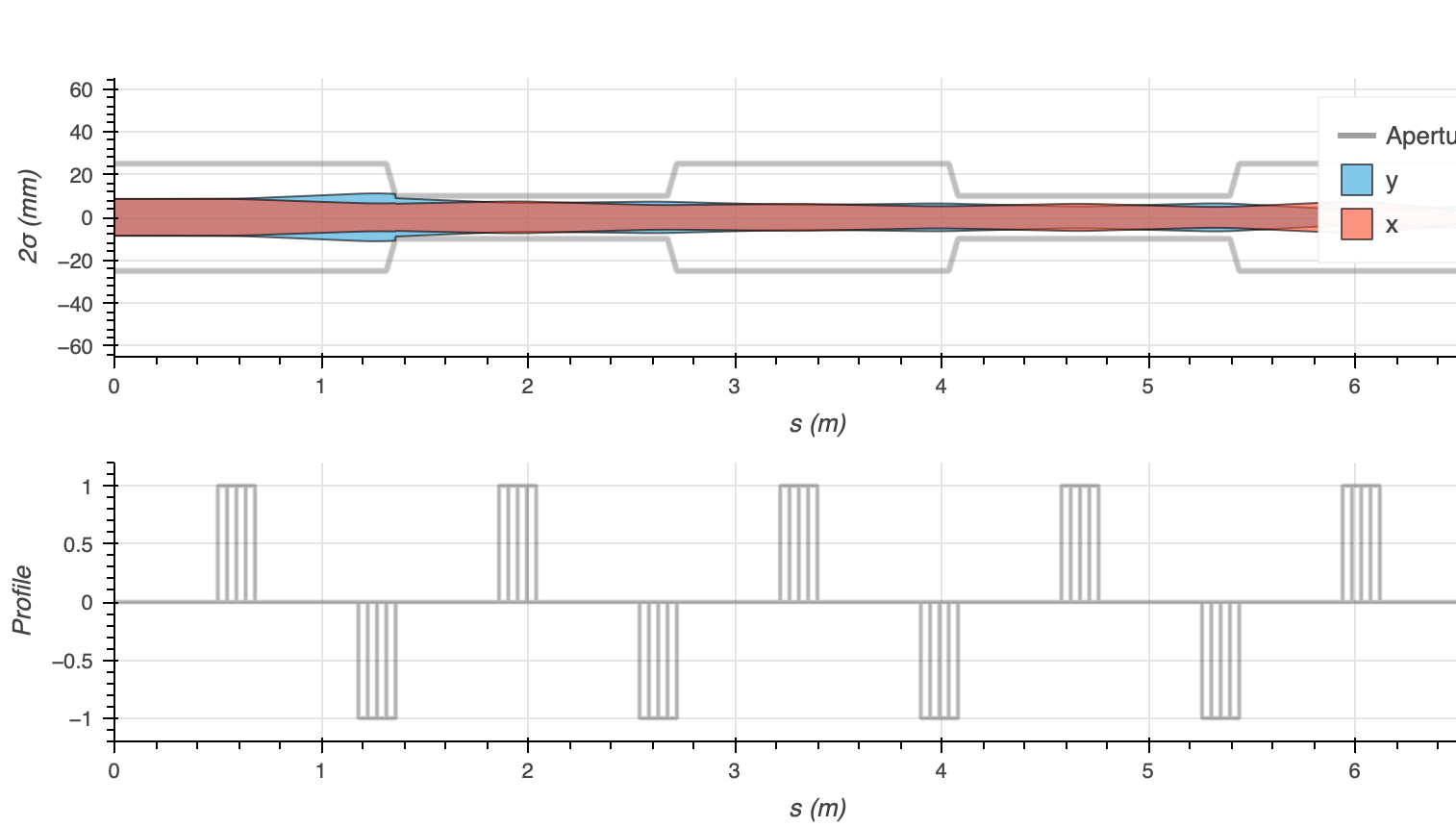}
    \caption{Beam envelope along the x-axis (red) and y-axis (blue) for the second beamline, with gray lines representing the aperture. The beam is effectively contained within the boundaries.}
    \label{fig:Envelope_Aperture_2}
\end{figure}

The second beamline addresses distinct challenges, including a varying aperture along its length and an elevated initial beam emittance of $2 \times 10^{-3}$ m·rad. Increased emittance leads to a more diffuse and less collimated beam, which can exacerbate particle losses, reduce luminosity in accelerator applications, and degrade overall performance. The changing aperture further complicates beam steering, as it requires precise adjustments to prevent scraping or loss at narrower sections. This beamline features 10 quadrupole magnets for beam control. Following training of our DDPG agent over 35,000 episodes, we attained a 91\% transmission rate. Figure~\ref{fig:charge_2} illustrates the beam charge progression, showing only minor reductions and thus low particle loss. Complementing this, Figure~\ref{fig:Envelope_Aperture_2} provides insights into the beam envelope along the x-axis (red) and y-axis (blue), with the gray aperture lines indicating that the beam is consistently confined within the boundaries, underscoring the agent's success in navigating these constraints.

\section{Discussion and Future Work}
Our RLABC framework showcases the promise of reinforcement learning in automating beamline optimization, significantly reducing the time and expertise required compared to traditional methods. By achieving transmission rates comparable to those obtained by physics experts on complex beamlines with bends, varying apertures, and elevated emittance, RLABC validates the effectiveness of modeling accelerator tuning as a sequential decision-making process. The integration with Elegant simulations ensures accurate modeling while maintaining accessibility through automated preprocessing and environment setup. However, the current implementation is limited to simulation-based optimization and relies on DDPG, which, while effective, may not generalize optimally across all beamline architectures due to variations in complexity and dynamics.
Looking ahead, we plan to expand RLABC by experimenting with additional RL algorithms, such as Proximal Policy Optimization (PPO) and Soft Actor-Critic (SAC), to identify the most robust approaches for diverse scenarios. We also intend to test the framework on a broader range of beamline architectures, to enhance its generalizability. Our main focus is the development of the algorithm, which is able to solve diverse beamline architectures without training from scratch for one particular environment. Furthermore, we envision evolving RLABC into a physicist's copilot system that provides real-time suggestions for beamline optimizations. For RL researchers, we aim to refine the library as a plug-and-play platform where custom agents can be seamlessly integrated and benchmarked against accelerator-specific challenges.

\section{Conclusion}

In this paper, we introduced RLABC, a Python library that leverages reinforcement learning to automate beamline optimization in particle accelerators using Elegant simulations. By framing the magnet tuning process as an RL task with well-defined states, actions, and rewards, and applying the DDPG algorithm, we attained high transmission efficiencies on challenging beamlines—results comparable to those achieved through expert manual tuning. RLABC modifies beamline optimization, making it accessible to non-experts while offering a flexible platform for advanced research. This work underscores the potential of RL in accelerator physics, providing a practical tool for further experimentation and establishing a foundation for future integration with real hardware and broader algorithmic explorations.

\bibliographystyle{ws-ijmpe}
\bibliography{sample}

\end{document}